\documentclass[11pt,a4paper]{article}

\usepackage{jcappub}
\newcommand{\beqa}{\begin{eqnarray}}
\newcommand{\eeqa}{\end{eqnarray}}
\newcommand{\mean}[1]{\ensuremath{\langle #1 \rangle}}

\title{Constraints on Ultracompact Minihalos from
Extragalactic $\gamma$-ray Background}

\author[a,b,c] {Yupeng Yang,}
\author[a,c] {Lei Feng,}
\author[b] {Xiaoyuan Huang,}
\author[b]{ Xuelei Chen,}
\author[c,d] {Tan Lu}
\author[a,c] {and Hongshi Zong}
\affiliation[a]{Department of Physics, Nanjing University, \\
Nanjing, 210093, China}
\affiliation[b]{National Astronomical Observatories, Chinese Academy of Sciences, \\
Beijing, 100012, China}
\affiliation[c]{Joint Center for Particle, Nuclear Physics and Cosmology, \\
Nanjing, 210093, China}
\affiliation[d]{Purple Mountain Observatory, Chinese Academy of Sciences, \\
Nanjing, 210008, China}
\emailAdd{yyp@chenwang.nju.edu.cn}
\emailAdd{fenglei@chenwang.nju.edu.cn}
\emailAdd{huangxiaoyuan@gmail.com} 
\emailAdd{xuelei@cosmology.bao.ac.cn}
\emailAdd{t.lu@pmo.ac.cn}
\emailAdd{zonghs@chenwang.nju.edu.cn}

\date{today}

\abstract{It has been proposed that ultracompact minihalos (UCMHs) 
might be formed in earlier epoch. If dark matter 
is in the form of Weakly Interacting Massive Particles (WIMPs), 
UCMHs can be treated as the $\gamma$-ray sources because of 
the dark matter annihilation within them. In this paper, 
we investigate the contributions of UCMHs formed during three 
phase transitions (electroweak symmetry breaking, QCD
confinement and $e^{+}e^{-}$ annihilation) to the extragalactic $\gamma$-ray background. 
Moreover, we use the Fermi-LAT observation data of the extragalactic 
$\gamma$-ray background to get the constraints on the 
current abundance of UCMHs produced during phase transitions.
We also compare these results with those obtained from Cosmic Microwave
Background (CMB) observations finding that the constraints from the Fermi-LAT 
are more stringent than CMB.}


\begin{document}

\maketitle
\section{Introduction}
It has been known that the present structures of cosmology 
originate from the density perturbations ($\delta \sim 10^{-5}$) 
in earlier epoch. Recently, the authors of \cite{0908.0735} 
have proposed that a new class of dark matter structure, 
ultracompact minihalos (UCMHs) would be formed if the density 
perturbations are between $3 \times 10^{-4}$ and 0.3 
which are not enough to produce the primordial black holes (PBHs).  
Although within the conventional cases the density perturbations are not larger
enough to form these objects, they could be enhanced
through the inflation potential or during 
the phase transitions in the early Universe \cite{PRL_1997}.
On the other hand, if dark matter is in the form of 
Weakly Interacting Massive Particles (WIMPs), UCMHs would become the sources
of $\gamma$-ray due to the dark matter annihilation within them. 
In paper \cite{scott_PRL_2009}, the authors have investigated 
the integrated $\gamma$-ray flux of nearby UCMHs formed during 
three transitions: electroweak symmetry breaking (EW), QCD
confinement and $e^{+}e^{-}$ annihilation. 
They find that for the same mass fraction, 
the flux from UCMHs formed during $e^{+}e^{-}$ within 
the 100pc would have been observed by EGRET or Fermi-LAT. 
The authors of \cite{josan_PRD_2010} have obtained the constraints on 
the abundance of UCMHs using the point source sensitivity 
of $Fermi$ above 100 MeV from the neighborhood: $f_{UCMHs} \sim 10^{-7}$ for 
$M_{UCMHs} \sim 10^{3} M_{\odot}$. The constraints on the UCMHs formed during 
$e^{+}e^{-}$ annihilation using 
the WMAP-7 years data and the forcast for the Planck-3 are obtained by authors 
in \cite{yyp,epjp}. 

\par
The extragalactic $\gamma$-ray background 
has been observed by $Fermi$ satellite \cite{fermi_extra} and 
its origination has not known exactly now. 
The main sources would be the usual, unresolved astrophysical objects, 
such as the Active Galactic Nuclei (AGN), normal galaxies and 
the clusters of galaxies \cite{dermer_2007,vasiliki_02}. 
Other possible sources may be the dark matter annihilation 
\cite{mack_94,elsosser_04,ullio_02}.
In our paper, we consider the contributions of dark matter annihilation 
from UCMHs, and get the constraints 
on the current abundance of them using the $Fermi$ observations. 

In addition, the annihilation of dark matter has an effect 
on the cosmological evolution \cite{lezhang_PRD_2006} e.g. recombination 
and reionization. So the property of dark matter 
can be constrained by the CMB observations. 
UCMHs would have the similar effect because of their 
higher density, and the CMB data can give constraints on their nature.  
Here, we also use the CMB data 
to investigate the current abundance of UCMHs.

\par

This paper is organized as following: 
we calculate the extragalactic $\gamma$-ray
background from UCMHs in section II. 
In section III, we get the constraints on 
the current abundance of UCMHs using the 
$Fermi$ and CMB data, and we conclude in section IV.

\section{Extragalactic $\gamma$-ray Background}
\subsection{Extragalactic $\gamma$-ray background from UCMHs}
After the formation of UCMHs, they can accret the dark matter 
particles onto them by radial infall due to their higher density. 
The mass of UCMHs changes as \cite{scott_PRL_2009}:

\begin{equation}
\label{Mh}
M_\mathrm{UCMHs}(z) = \delta m \left(\frac{1 + z_\mathrm{eq}}{1+z}\right),
\end{equation}

where the $\delta m$ is the mass contained within a perturbation 
at the redshift of matter-radiation equality $z_\mathrm{eq}$. 
Following \cite{scott_PRL_2009}, we adopt the value of 
$\delta m = 5.6 \times 10^{-19} M_{\odot}, 1.1 \times 10^{-9} M_{\odot}, 0.33 M_{\odot}$ 
for three phase transitions: EW, QCD and $e^{+}e^{-}$.

\par

The density profile of UCMHs is \cite{scott_PRL_2009}:
\begin{equation}
\label{density}
\rho_{UCMHs}(r,z) = \frac{3f_\chi M_\mathrm{UCMHs}(z)}{16\pi R_\mathrm{UCMHs}(z)^\frac{3}{4}r^\frac{9}{4}},
\end{equation}

here ${R_\mathrm{UCMHs}(z)} = 
0.019\left(\frac{1000}{z+1}\right)\left(\frac{M_\mathrm{UCMHs}(z)}
{\mathrm{M}_\odot}\right)^\frac{1}{3} \mathrm{pc}$ 
and $f_{\chi} = \frac{\Omega_{DM}}{\Omega_b+\Omega_{DM}} = 0.83$ \cite{wmap} 
is the dark matter fraction.
We accept the assumption that UCMHs 
stop growing at $z \approx 10$ because 
the structure formation process prevents further accretion 
after the redshift. 

\par

\begin{figure}
\epsfig{file=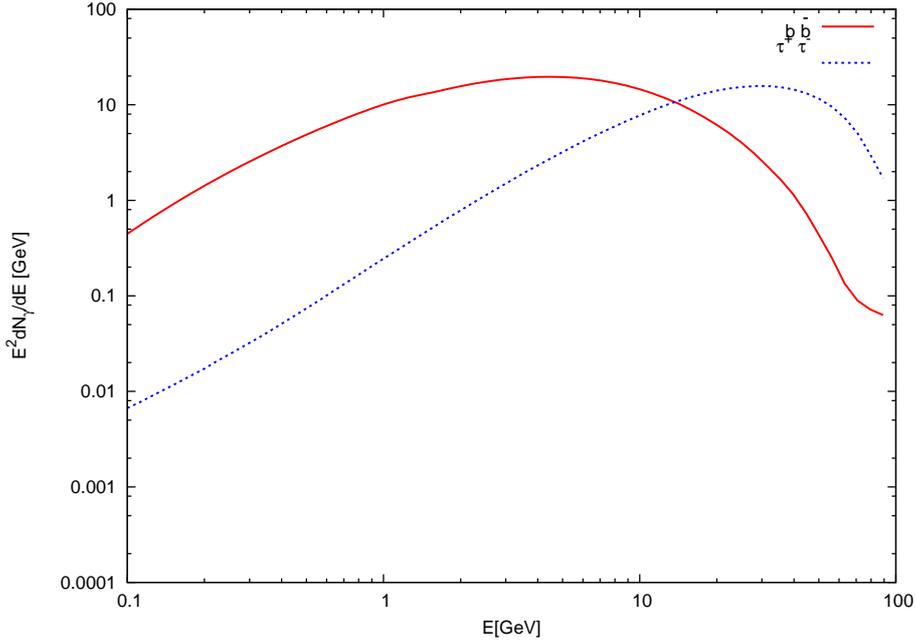,width=0.8\columnwidth}
\caption{The annihilation channels ($\tau^{+}\tau^{-}$ 
and $b \overline b$) used in this work are shown. Here we have choosed 
the mass of dark matter $M_{\chi} = 100 GeV$.}
\label{fig:example}
\end{figure}

The diffuse extragalactic $\gamma$-ray flux from 
the annihilation of dark matter can be written as \cite{ullio_02}:

\beqa
  \frac{d\phi_{\gamma}}{dE} 
   =  \frac{c}{4 \pi} \int^{z_{up}}_0 dz \frac{e^{-\tau(z,E_0)}}{H(z)} \int dM \frac{dn}{dM}(M,z)
  \frac{d{\cal N}_{\gamma}}{dE}\left(E_0\,(1+z),z\right)\;. 
\label{eq:flux1}
\eeqa

where $H(z)=H_{0}\sqrt{\Omega_M(1+z)^3+\Omega_\Lambda}$. 
The upper limitation of integration $z_{up} = M_{\chi}/E_0 - 1$ \cite{0105048}.
$\tau$ is the optical depth, and three processes 
are considered for our purpose \cite{1989_apj}: 
(i) photon-matter pair production, (ii) photon-photon scattering, 
(iii) photon-photon pair production. In fact, as shown in 
\cite{1989_apj,xuelei_PRD_2004,0906.1197}, the energy range 
($10^{8} eV \lesssim E \lesssim10^{11} eV$) and 
redshift of $\gamma$-ray ($0 \lesssim z \lesssim 1100$)   
in which we are interest are not affected by the medium. It is corresponding to 
the 'Transparency window' of photons, and these photons will propogate 
freely almost. For the energy range of photons $E \lesssim 10^{5} eV$
and $E \gtrsim 10^{11} eV$, 
there are stronger absorption by the medium during all redshift 
and dot not form the $\gamma$-ray background. 
${dn}/{dM}$ is the mass function. In our paper,
we consider the monochromatic mass fuction for UCMHs, which means 
all of them have same mass \cite{pbhs}: 
${dn}/{dM}(M,z) \sim \delta(M-M(t_{i}))$. 
We assume that the abundance of UCMHs is same everywhere 
and they do not merger with others. 
$d{\cal N}_{\gamma}/{dE}$ is the number of $\gamma$-rays 
from one of UCMHs per unit of time and energy: 
  
\beqa
  \frac{d{\cal N}_{\gamma}}{dE} (E,z) & = & \frac{\mean{\sigma v}}{2}
  \frac{dN_{\gamma}(E)}{dE} B_{f} 
  \int \left(\frac{\rho_{UCMHs}(r,z)}{M_{\chi}}\right)^2 d^3r
\label{eq:dnde}
\eeqa
$\mean{\sigma v}$ is the thermally cross section of dark matter 
and $M_{\chi}$ is the mass of dark matter.  
$dN_{\gamma}/dE$ is the energy spectrum of the $\gamma$-ray 
from dark matter annihilation. Here, we only consider the prompt emission
case and use the public code DarkSUSY \cite{darksusy} 
to calculate for different channels. In this work, we consider two typical 
channels $\tau^{+}\tau^{-}$ and $b \overline b$. 
In Fig:~\ref{fig:example}, the annihilation spectrum of 
these two channels are shown.

$B_{f}$ is the branch ratio of annihilation. In reality, 
different channels corresponds to different branching ratios.
Here we consider each channel separately and take $B_{f}$ = 1. 

Base on these assumptions, we can write the differential energy spectrum 
of extragalactic $\gamma$-rays from UCMHs: 


\beqa
\left.  \frac{d\phi_{\gamma}}{dE} \right |_{\rm UCMHs}
   = n_{UCMHs,0}\frac{c}{8 \pi} \frac{\mean{\sigma v}}{M_{\chi}^2} \int dz \frac{(1+z)^3 e^{-\tau(E,z)}}{H(z)}
  \frac{dN_{\gamma}}{dE}\left(E,z\right) \int \rho_{UCMHs}^2(r,z) d^3r \nonumber \\
   = \frac{f_{UCMHs,0} \rho_{c,0}}{M_{UCMHs,0}}\frac{c}{8 \pi} 
\frac{\mean{\sigma v}}{M_{\chi}^2} \int^{z_{up}}_0 dz \frac{(1+z)^3 
e^{-\tau(E,z)}}{H(z)}\frac{dN_{\gamma}}{dE}\left(E,z\right) \int_{r_{cut}}^{R} 
\rho_{UCMHs}^2(r,z) d^3r
\label{eq:flux}
\eeqa

$n_{UCMHs,0}$ is the number density now, $f_{UCMHs,0}$ 
is the current abundance of UCMHs and 
defined as $f_{UCMHs} = \rho_{UCMHs}/\rho_{c}$, 
$\rho_{c,0}$ is the critical density at present
and $M_{UCMHs,0}$ is the mass 
of UCMHs at $z = 0$. Due to the annihilation of dark matter, there is 
the maximum density at present for a halo, so the $r_{cut}$ can be defined as \cite{0509565,0207125}: 

\begin{equation}
\rho(r_{cut}) = \rho_{max} = \frac{M_{\chi}}{\mean{\sigma v}(t_{0}-t_{i})}
\end{equation}

where $t_{0} = 13.7 Gyr$ is the age of universe, $t_{i}$ is the time 
of UCMHs formation and we adopt $t_{i}(z_{eq}) = 77kyr$ used by \cite{josan_PRD_2010}. During the radiation dominated era, 
the growth of the UCMHs is very slow until the equality of matter and radiation.
So we use the same value of $t_i$ for three cases.
Following \cite{scott_PRL_2009,josan_PRD_2010}, 
we assume that the density is constant within 
$r_{cut}$, $\rho(r \le r_{cut}) = \rho(r_{cut})$.

\subsection{Extragalactic $\gamma$-ray background from halos}
 
Besides the UCMHs, we include the extragalactic $\gamma$-ray 
background from halos, which is also produced 
by dark matter annihilation \cite{ullio_02} \cite{0105048}: 

\begin{eqnarray}
\label{eq:fai}
\left. \frac{d\phi_\gamma}{dE} \right |_{\rm Halos}=
\frac{c}{8 \pi} \frac{{\rho}_{c,0}^2\mean{\sigma v}}{M_{\chi}^2}
\int^{z_{up}}_0 {dz (1+z)^3 \frac{C(z)}{H(z)} 
\frac{dN_\gamma}{dE}(E,z)B_{f}e^{-\tau(E,z)}}
\end{eqnarray}

here $C(z)$ is the 'clumping factor'  
relative to the homogeneous case \cite{cumberbatch}, 
while the process of structure formation is considered.

\begin{eqnarray}
\label{eq:cz}
C(z)&=&1 + {\Gamma_{halo}(z) \over \Gamma_{smooth}(z)}\nonumber\\
&&=1 + {(1 + z)^3 \over \bar \rho_{\rm DM}^2(z)}\int 
{\rm d}M\frac{{\rm d}n}{{\rm d}M}(M,z) \\
&&\times\int\rho^2(r)4\pi r^2{\rm d}r\nonumber 
\end{eqnarray}

where $\Gamma_{s}$ stand for the dark matter annihilation rate. 
$dn/dM$ is the halos mass function 
and we use the Press-Schechters formalism \cite{ps}.  
On the other hand, it has been found that  
there are many substructures in dark matter halos \cite{diemand}. 
These sub-halos can also enhance the dark matter 
annihilation rate. In our paper, we  
include these sub-halos, 
while neglect the contributions 
from the sub-sub-halos and 
use the smallest mass of them 
$\sim 10^{-6}M_\odot$ \cite{green,diemand}.  
We consider about $\sim 10\%$ halos 
mass within the sub-halos, 
use the power low 
form of mass function $\sim M^{-\beta}$ and adopt 
$\beta = 1.95$ \cite{diemand}. 
So the total clumping factor of dark matter 
halos and sub-halos can be written as \cite{cumberbatch}: 

\begin{equation}
C_{total} = 1 + (C_{halos} - 1) + (C_{subhalos} - 1)
\end{equation}

\begin{figure}
\epsfig{file=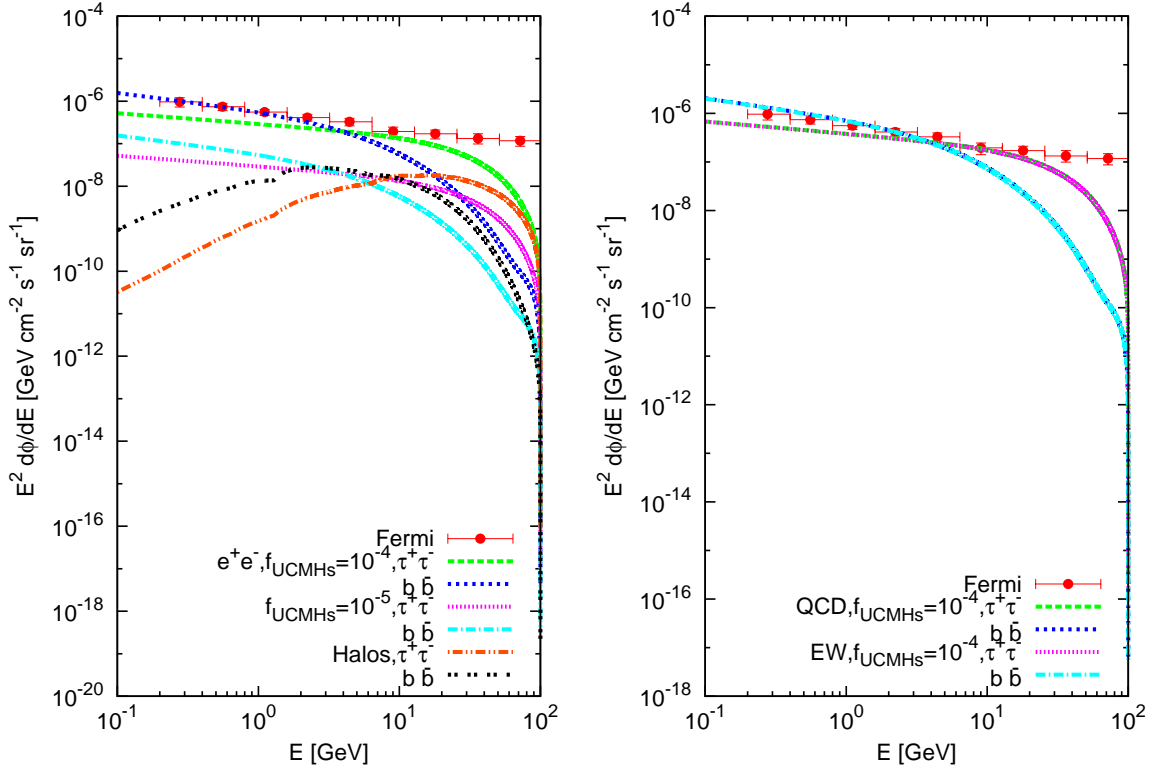,width=\columnwidth}
\caption{The extragalactic $\gamma$-ray background from UCMHs for different
phase transitions, channels and current abundance. 
The contributions of halos are also shown. Here we have set 
$\mean{\sigma v}$ = $3 \times 10^{-26} cm^{3}s^{-1}$ and 
$M_{\chi} = 100 GeV$, and for the other cosmological 
parameters we use the WMAP results \cite{wmap}.
The $Fermi$ data are obtained from \cite{fermi_extra}.}
\label{fig:gamma}
\end{figure}

\begin{figure}
\epsfig{file=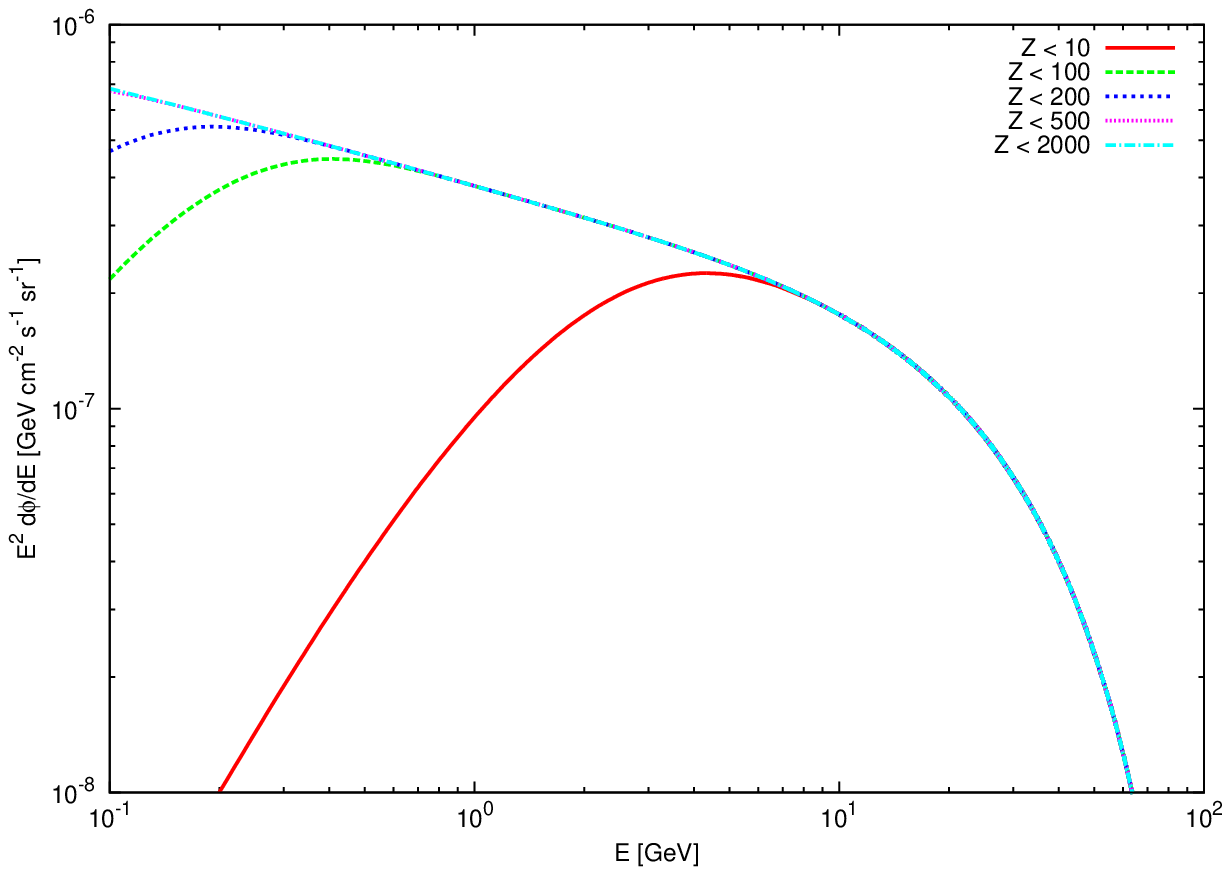,width=0.8\columnwidth}
\caption{The extragalactic $\gamma$-ray background from UCMHs formed 
during the $e^{+}e^{-}$ annihilation for different
redshift ranges, where the $\tau^{+}\tau^{-}$ channel is shown. Here we have set 
$f_{UCMHs} = 10^{-4}$, $\mean{\sigma v}$ = $3 \times 10^{-26} cm^{3}s^{-1}$ and 
$M_{\chi} = 100 GeV$, and for the other cosmological 
parameters we use the WMAP results \cite{wmap}.}
\label{fig:gfz}
\end{figure}

In Fig.~\ref{fig:gamma}, for two typical channels $\tau^{+}\tau^{-}$, $b\bar b$,
we show the extragalactic $\gamma$-ray background 
from UCMHs formed during EW, QCD, $e^{+}e^{-}$ phase transitions, 
where the $Fermi$ data are from \cite{fermi_extra}. 
Here we have set $\mean{\sigma v} = 
3.0 \times 10^{-26} cm^{3} s^{-1}$ and $M_{\chi} = 100 GeV$.
For the cosmological parameters we use the WMAP results \cite{wmap}.
For the case $f_{UCMHs}  = 10^{-4}$, the flux approaches 
the $Fermi$ observation data especially for lower energy. 
The flux from those objects formed during 
the QCD and EW is same almost.
We also show the $\gamma$-ray background from the halos. 
In Fig:~\ref{fig:gfz}, The $\gamma$-ray background from UCMHs for different redshift
are shown. From there we can see that the main contributions come from 
$z \lesssim 200$.

\section{Constraints on the current abundance of UCMHs }
The production of dark matter annihilation 
would be standard particles, such as photons, 
electrons and positrons. These particles 
have interaction with the cosmological medium. 
So the processes of recombination and reionization  
can be influenced by the dark matter annihilation \cite{lezhang_PRD_2006}. 
It is similar for UCMHs because of their higher density which 
can cause larger annihilation rate. The annihilation rate 
of UCMHs can be written as: 

\begin{eqnarray}
\label{eq:fai}
\Gamma = n_{UCMHs} \Gamma' = n_{UCMHs} \frac{\langle\sigma v\rangle}{M_\chi^2} \int 4\pi r^2\rho^2(r,z)\mathrm{d}r \nonumber\\
= \frac{\rho_{UCMHs,0}}{M_{UCMHs,0}}(1+z)^3 
\frac{\langle\sigma v\rangle}{M_\chi^2} \int 4\pi r^2\rho^2(r,z)\mathrm{d}r \nonumber\\
= \frac{f_{UCMHs,0} \rho_{c,0}}{M_{UCMHs,0}}(1+z)^3 
\frac{\langle\sigma v\rangle}{M_\chi^2} \int 4\pi r^2\rho^2(r,z)\mathrm{d}r.
\end{eqnarray}

here $\Gamma'$ is the annihilation rate within one of UCMHs and 
$\Gamma$ is the annihilation rate per unit volume of UCMHs.

Considering the dark matter annihilation, 
the evolution of ionization fraction $x_{e}$ 
can be written as \cite{lezhang_PRD_2006,xuelei_PRD_2004,0905.0003}:
\begin{equation}
(1+z){dx_e\over dz} = {1\over H(z)}[R_s(z)-I_s(z)-I_\chi(z)]
\end{equation}
where $R_s$ is the standard recombination rate, 
$I_s$ is the ionization rate by standard sources, 
$I_\chi$ is the ionization rate sourced by dark matter 
which is given as \cite{lezhang_PRD_2006}: 
\begin{equation}
{I_\chi} = {\chi_i f{2 m_\chi c^{2}\over n_b E_b}}{\Gamma_{total}} 
\end{equation}
where $n_b$ is the baryon number density and the $E_b = 13.6 eV$ 
is the ionization energy.
$\Gamma_{total}$ is the total dark matter
annihilation rate including the UCMHs and halos. 
$f$ that depends on the redshift and  
the production of dark matter annihilation \cite{slatyer}
is the released energy fraction 
depositing in the baryonic gas
during the annihilation. 
In our paper, we assume that the 
total energy released by the 
annihilation is deposited, which means $f = 1$.
$\chi_i$ is the energy fraction which 
ionizes the baryonic gas and we  
accept the form given by \cite{xuelei_PRD_2004}: 
\begin{equation}
{\chi_i} = {\left(1 - {x_e}\right) / 3}
\end{equation}  
here $x_e$ is the fraction of free electrons.

\begin{figure}
\epsfig{file=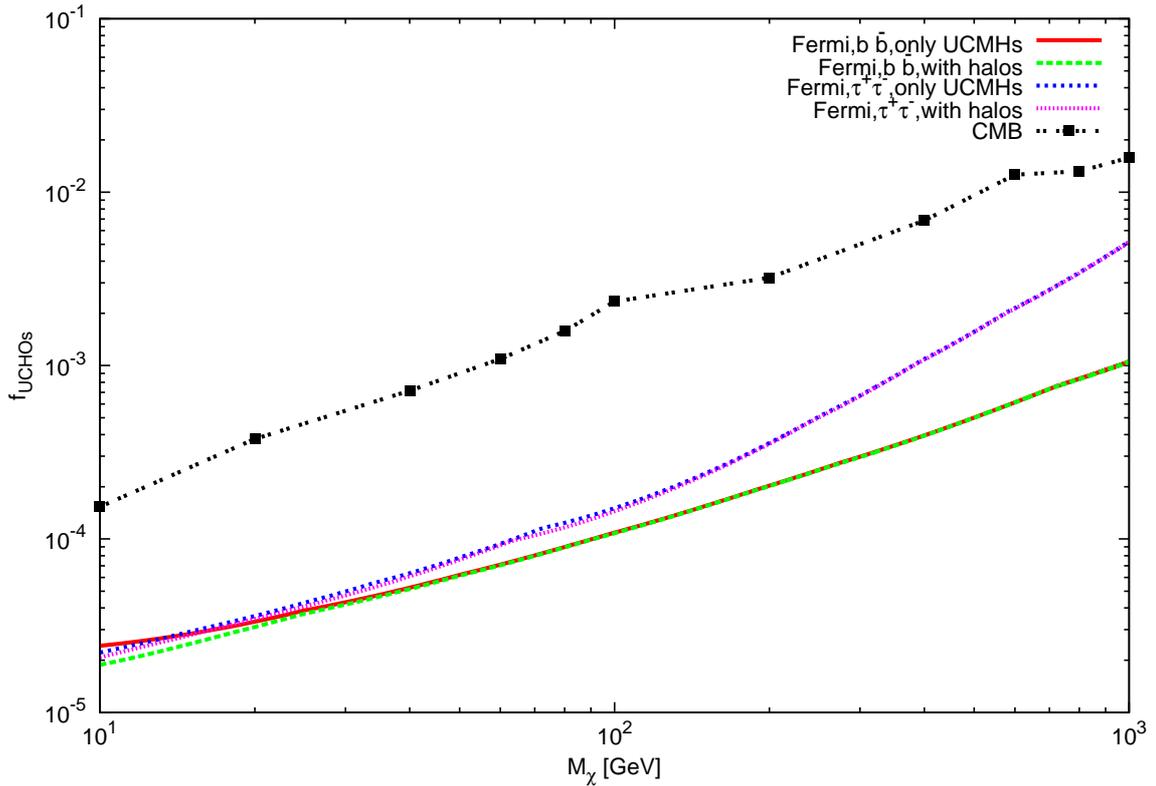,width=\textwidth}
\caption{The constraints on the current abundance 
of UCMHs produced during $e^{+}e^{-}$: 
$f_{UCMHs}$ from $Fermi$ and CMB observations. 
The model and parameters used have been described in text. 
The constraints using the CMB data are obtained from Tab. 1.}
\label{fig:com}
\end{figure}

We have modified the public code CAMB \cite{camb} 
and COSMOMC \cite{cosmomc} in order to 
include the effect of UCMHs, halos and get the constraints on 
the parameters. We use the CMB data including 
seven years WMAP data \cite{wmap}, 
and the data from ACBAR \cite{acbar}, Boomerang \cite{boomerang}, 
CBI \cite{cbi} and VSA \cite{vsa} experiments.
We consider 6 cosmological parameters:
$\Omega_b h^2,\Omega_d h^2,\theta,\tau,n_s,A_s$,
where $\Omega_b h^2$ and $\Omega_d h^2$  are the density of
baryon and dark matter, $\theta$ is the ratio of
the sound horizon at recombination to its angular diameter distance
multiplied by 100, $\tau$ is the optical depth,
$n_s$ and $A_s$ are the spectral index and amplitude of
the primordial density perturbation power spectrum. 
We fix the value of $\mean{\sigma v} = 3 \times 10^{-26} cm^3 s^{-1}$ 
and treat the current abundance of UCMHs ($f_{UCMHs}$) and 
the mass of dark matter ($M_{\chi}$) as the free parameters. 
As it has been shown that the contributions 
of UCMHs formed during three phase transitions 
are same almost, so here we consider the $e^{+}e^{-}$ 
case simply.
We choose the mass of dark matter: 
1, 10, 20, 40, 60, 80, 100, 200, 400, 600, 800, 1000 GeV and get 
the corresponding $2\sigma$ value of $f_{UCMHs}$ respectively.  
The results are shown in Tab.~\ref{tab:ucmhs} and Fig.~\ref{fig:com}.
From these we can see that for the larger dark matter mass, 
the allowed value of $f_{UCMHS}$ becomes larger gradually. 
The similar discussion and same method are also present in \cite{yyp}, 
where the authors change the mass of dark matter for MCMC method and get 
the constraints on the UCMHs. The results are consistent between of them.
On the other hand, for the constrains on $f_{UCMHs}$ from CMB, 
the main contribution comes from the recombination \cite{epjp}. During the 
reionization period, the standard sources such as stars or active galactic
nuclei (AGN) are dominant. So the including of common dark matter halos 
is not so important, this is shown in \cite{epjp}, where the authors 
only consider the contribution from UCMHs and find the comparable results. 
As we have mentioned above, we have assumed that the energy released by dark matter 
annihilation is all deposited in the medium. It is not always the truth. 
For example, the neutrino can take the energy off without any effect 
on the medium. In the higher redshift, $z \sim 1000$, 
the fraction is about $30\% \sim 40\%$. It will decrease to 
$\sim 1\%$ in the lower redshift $z \sim 6$ \cite{0808.0881}. 
So if one consider these effect, the final results can 
be different for the different annihilation channels and productions \cite{0906.1197,1106.1528}. 
\footnote{Actually in our work, from Eq.(3.1) nad (3.3), we can also think that the two parameters 
used by us is the combination of $f$ and other parameters: $ff_{UCMHs}$ and $f^{-1}M_{\chi}$.
So the final results are the constraints of them.
For the different $f$ we can get the corresponding constraints on $f_{UCMHs}$ and $M_{\chi}$. 
Of course, in this case, we also treat $f$ as a constant instead of 
a function of redshift.}




\begin{table*}
\begin{center}
\begin{tabular}{cccccccc}
\hline
\hline
$M_{\chi}(GeV)$&$f_{UCMHs}$&$M_{\chi}(GeV)$&$f_{UCMHs}$&$M_{\chi}(GeV)$&$f_{UCMHs}$ \\
\hline
\\
1&$0.22 \times 10^{-4}$&60&$0.11 \times 10^{-2}$&400&$0.69 \times 10^{-2}$ \\
10&$0.15 \times 10^{-3}$&80&$0.16 \times 10^{-2}$&600&$0.12 \times 10^{-1}$ \\
20&$0.38 \times 10^{-3}$&100&$0.23 \times 10^{-2}$&800&$0.13 \times 10^{-1}$ \\
40&$0.72 \times 10^{-3}$&200&$0.32 \times 10^{-2}$&1000&$0.16 \times 10^{-1}$ \\
\hline
\hline
\end{tabular}
\caption{\label{tab:ucmhs}
 Posterior constraints on the current fraction of UCMHs 
formed during $e^{+}e^{-}$
for the different dark matter mass, where the $2\sigma$ values are shown.}
\end{center}
\end{table*}

As we have shown in section II, the extragalactic $\gamma$-ray 
flux from UCMHs formed during $e^{+}e^{-}$ phase transition 
would exceed the $Fermi$ observations for some parameters. 
In order to be consistent with $Fermi$ data, 
we get the $2\sigma$ conservative limitations of the current abundance of UCMHs 
using these data following the method in \cite{fermi_cons}:

\begin{eqnarray}
\phi^{\gamma}_{i} \leq M_{i}+n \times \Sigma_{i}
\end{eqnarray}

where $\phi^{\gamma}_{i}$ is the integrated flux from our model in $i$th 
eneregy bin corresponding to the measured flux, $M_i$. $\Sigma_i$ 
is the error of $i$th bin. $n = 1.28$ and 1.64 correspond to the $90\%$ and 
$95\%$ confidence level.  

The results are shown in Fig.~\ref{fig:com} 
where we have shown the $\tau^{+} \tau^{-}$, $b \bar b$ channels. 
We can see that the constraints for the $b \bar b$ channel 
are more stringent than those for the $\tau^{+}\tau^{-}$ channel especially 
for larger dark matter mass. 
Moreover, comparing with the results obtained from CMB, the constraints from 
$Fermi$ are much better. So the observations of extragalactic $\gamma$-ray 
background can give more stringent constraints on the 
current abundance of UCMHs than the CMB data. 
In Fig.~\ref{fig:com}, we also show the constraints including 
the contributions of halos.
\footnote{In this paper, we do not include the contributions
from the standard astrophysical sources.
If these objects are also considered,
then the allowed abundance of UCMHs
would be more lower. We will consider
these in the other papers.}
From the results we can see that the constraints are same nearly 
for with and without halos. There are some light 
differences for the $b \overline b$ channel 
in the lower dark matter mass $M_{\chi} \lesssim 30 $GeV.

\section{Conclusion}
We have investigated the extragalactic $\gamma$-ray background from 
a new class of dark matter structure named UCMHs and 
gotten the constraints on the current abundance of them using the $Fermi$ 
observations. On the other hand, 
because the dark matter annihilation within the UCMHs 
has effect on the evolution of cosmological ionization fraction, 
we also use the CMB data to get the constraints on the fraction of UCMHs. 
In our paper, we only show the constraints on the UCMHs formed during 
$e^{+}e^{-}$ due to the same contributions of UCMHs produced during EW and QCD.
\par
From the results, we find that the constraints 
are weaker for CMB data than $Fermi$ data. 
About the dark matter mass $M_{\chi} \sim 10 GeV$, the current fractions 
of UCMHs are about $\sim 10^{-5}$ and $\sim 10^{-4}$ respectively 
corresponding to $Fermi$ and CMB data. 
For the $Fermi$ observations, we have calculated 
two channels: $\tau^{+} \tau^{-}$ and $b \bar b$. We find that 
for lower mass of dark matter $M_{\chi} \lesssim 100 GeV$, 
the constraints are comparable, while for larger one the constraints 
from $b \bar b$ channel are more stringent. 
If we consider the contributions from the halos and subhalos, 
the results are not changed nearly especially for 
larger dark matter mass $M_{\chi} \gtrsim 100 GeV$. 
\par

For this new dark matter structures, it is hope 
that other present or the future observations can have 
the power to give new constraints on them. 
For the recent observations of positron, such as PAMELA \cite{pamela}, Fermi \cite{fermi} 
and ATIC \cite{atic}, may be the signal from the dark matter 
annihilation (for a review one can see \cite{0909.1182}). 
The characteristic of UCMHs would explain 
these observations. On the other hand, the extra energy injection 
into the medium can also have effect on the 21cm signals. 
So we expect that this future observations can also 
give the new and stronger constraints on the UCMHs.

Moreover, the constraints on the abundance of UCMHs can be translated 
into constraints on the primordial curvature perturbation~\cite{1110.2484}. We will consider 
them in the near future.


\acknowledgments
We would like to thank Weimin Sun and Yan Qin for
improving the manuscript.
Our MCMC chains computation is performed on
the Shenteng 7000 system of the Supercomputing
Center of Chinese Academy of Sciences. 
This work is supported in part by the National Natural Science Foundation of China 
(under GrantNos 10775069, 10935001 and 11075075) and the Research
Fund for the Doctoral Program of Higher Education 
(under Grant No 200802840009)


\begin{thebibliography}{10}

\bibitem{0908.0735}
M.~{Ricotti}, A.~{Gould}, 
  {\it A New Probe of Dark Matter and High-Energy Universe Using Microlensing}, 
  [\href{http://xxx.lanl.gov/abs/astro-ph/0908.0735}{{\tt 0908.0735}}].

\bibitem{PRL_1997}
C.~{Schmid}, D.~J. {Schwarz}, P.~{Widerin},
{\it Peaks above the Harrison-Zel’dovich Spectrum due to the Quark-Gluon
o Hadron Transition},  {\em Phys. Rev. Lett.} {\bf 78} (1997) 791-794,
  [\href{http://xxx.lanl.gov/abs/astro-ph/9806125}{{\tt astro-ph/9806125}}].

\bibitem{scott_PRL_2009}
P.~{Scott}, S.~{Sivertsson},
{\it Gamma Rays from Ultracompact Primordial Dark Matter Minihalos},  
{\em Phys. Rev. Lett.} {\bf 103} (2009) 211301,
  [\href{http://xxx.lanl.gov/abs/0908.4082}{{\tt 0908.4082}}].

\bibitem{josan_PRD_2010}
A.~S. {Josan}, A.~M. {Green}, 
{\it Gamma-rays from ultracompact minihalos: 
potential constraints on the primordial curvature perturbation},
{\em Phys. Rev.} {\bf D82} (2010) 083527,
  [\href{http://xxx.lanl.gov/abs/1006.4970}{{\tt 1006.4970}}].


\bibitem{yyp}
Y.~{Yang}, X.~{Huang}, X.~{Chen}, H.~{Zong}, 
{\it New Constraints on Primordial Minihalo Abundance Using
Cosmic Microwave Background Observations },
{\em Phys. Rev.} {\bf D84} (2011) 043506,
  [\href{http://xxx.lanl.gov/abs/1109.0156}{{\tt 1109.0156}}]

\bibitem{epjp}
Y.~{Yang}, X.~{Chen}, T.~{Lu}, H.~{Zong}, 
{\it The Abundance of New Kind of Dark Matter Structures},
Eur. Phys. J. PLUS 126(2011)123.

\bibitem{fermi_extra}
{\bf Fermi-LAT} Collaboration, A.~A. Abdo {\em et~al.},
{\it The Spectrum of the Isotropic Diffuse Gamma-Ray Emission 
Derived From First-Year Fermi Large Area Telescope Data},
{\em Phys. Rev. Lett.} {\bf 104} (2010) 101101,
  [\href{http://xxx.lanl.gov/abs/1002.3603}{{\tt 1002.3603}}].










\bibitem{dermer_2007}
C.~D. {Dermer},
{\it The Extragalactic Gamma Ray Background},
{\em AIP. Conf. Proc.} {\bf 921} (2007) 122-126,
  [\href{http://xxx.lanl.gov/abs/0704.2888}{{\tt 0704.2888}}].

\bibitem{vasiliki_02}
V.~{Pavlidou}, B.~D. {Fields},  
{\it The Guaranteed Gamma-Ray Background },
{\em Astrophys. J.} {\bf 575} (2002) L5-L8,
  [\href{http://xxx.lanl.gov/abs/astro-ph/0207253}{{\tt astro-ph/0207325}}].

\bibitem{mack_94}
M.~{Kamionkowski}, 
{\it Diffuse Gamma-Ray from WIMP Decay and Annihilation},
  [\href{http://xxx.lanl.gov/abs/astro-ph/9404079}{{\tt astro-ph/9407079}}].

\bibitem{elsosser_04}
D.~{Elsosser}, K.~{Mannheim}, 
{\it Supersymmetric Dark Matter and the Extragalactic Gamma Ray Background },
{\em Phys. Rev. Lett.} {\bf 94} (2005) 171302,
  [\href{http://xxx.lanl.gov/abs/astro-ph/0405235}{{\tt astro-ph/0405235}}].




\bibitem{ullio_02}
P.~{Ullio}, L.~{Bergstrom}, J.~{Edsjo}, C.~{Lacey}, 
{\it Cosmological dark matter annihilations into gamma-rays - a clock look },
{\em Phys. Rev.} {\bf D66} (2002) 123502,
  [\href{http://xxx.lanl.gov/abs/astro-ph/0207125}{{\tt astro-ph/0207125}}].

\bibitem{lezhang_PRD_2006}
L.~{Zhang}, X.~L. {Chen}, Y.~A. {Lei}, Z.~G. {Si},
{\it The impacts of dark matter particle annihilation 
on recombination and the anisotropies of the cosmic microwave background },
{\em Phys. Rev.} {\bf D74} (2006) 103519,
  [\href{http://xxx.lanl.gov/abs/astro-ph/0603425}{{\tt astro-ph/0603425}}].

\bibitem{0905.0003}
S.~{Galli}, F.~{Iocco}, G.~{Bertone}, A.~{Melchiorri}, 
{\it CMB constraints on Dark Matter models with large annihilation cross-section },
{\em Phys. Rev.} {\bf D80} (2010) 023505,
  [\href{http://xxx.lanl.gov/abs/0905.0003}{{\tt 0905.0003}}].



\bibitem{wmap}
E.~{Komatsu} {\em et~al.},
{\it Seven-Year Wilkinson Microwave Anisotropy Probe (WMAP)
Observations: Cosmological Interpretation},
{\em Astrophys,J,Suppl.} {\bf 192} (2011) 18,
[\href{http://xxx.lanl.gov/abs/1001.4538}{{\tt 1001.4538}}].


\bibitem{0105048}
L.~{Bergstrom}, J.~{Edsjo}, P.~{Ullio},
{\it Spectral Gamma-ray Signatures of Cosmological Dark Matter Annihilations},
{\em Phys. Rev. Lett.} {\bf 87} (2001) 251301,
  [\href{http://xxx.lanl.gov/astro-ph/0105048}{{\tt astro-ph/0105048}}].

\bibitem{1989_apj}
A.~A. {Zdziarski}, R.~{Svensson}
{\it Absorption of X-rays and Gamma Rays at Cosmological Distances},
{\em Astrophys. J.} {\bf 344} (1989) 551-566.

\bibitem{0906.1197}
T.~{Slatyer}, N.~{Padmanabhan}, D.~{Finkbeiner}, 
{\it CMB Constraints on WIMP Annihilation: Energy Absorption During the Recombination Epoch},
{\em Phys. Rev.} {\bf D80} (2009) 043526, 
[\href{http://xxx.lanl.gov/abs/0906.1197}{{\tt 0906.1197}}].

\bibitem{xuelei_PRD_2004}
X.~{Chen}, M.~{Kamionkowski}, 
{\it Particle Decays During the Cosmic Dark Ages},
{\em Phys. Rev.} {\bf D70} (2004) 043502, 
[\href{http://xxx.lanl.gov/abs/astro-ph/0310473}{{\tt astro-ph/0310473}}].

\bibitem{pbhs}
B.~J. {Carr}, K.~{Kohri}, Y.~{Sendouda}, J.~{Yokoyama}, 
{\it New cosmological constraints on primordial black holes },
{\em Phys. Rev.} {\bf D81} (2010) 104019,
  [\href{http://xxx.lanl.gov/abs/0912.5297}{{\tt 0912.5297}}].

\bibitem{darksusy}
http://www.physto.se/~edsjo/darksusy/

\bibitem{0509565}
G.~{Bertone}, A.~{Zentner}, J.~{Silk}, 
{\it A New Signature of Dark Matter Annihilations: Gamma-Rays from Intermediate-Mass Black Holes},
{\em Phys. Rev.} {\bf D72} (2005) 103517, 
[\href{http://xxx.lanl.gov/abs/astro-ph/0509565}{{\tt astro-ph/0509565}}].

\bibitem{0207125}
P.~{Ullio}, L.~{Bergstrom}, J.~{Edsjo}, C.~{Lacey}, 
{\it Cosmological dark matter annihilations into gamma-rays - a closer look },
{\em Phys. Rev.} {\bf D66} (2002) 123502, 
[\href{http://xxx.lanl.gov/abs/astro-ph/0207125}{{\tt astro-ph/0207125}}].






\bibitem{cumberbatch}
D.~T. {Cumberbatch}, M.~{Lattanzi}, J.~{Silk}, 
{\it Signatures of clumpy dark matter in the global 21 cm background signal},
{\em Phys. Rev.} {\bf D82} (2010) 103508,
  [\href{http://xxx.lanl.gov/abs/0808.0881}{{\tt 0808.0881}}].

\bibitem{ps}
W.~H. {Press}, P.~J. {Schechter}, 
{\it Formation of Galaxies and Clusters of Galaxies by Self-Similar Gravitational Condensation},
{\em Astrophys. J.} {\bf 187} (1974) 425-438.

\bibitem{diemand}
J.~{Diemand}, B.~{Moore}, J.~{Stadel}, 
{\it Earth-mass dark-matter haloes as the first structures in the early Universe},
{\em Nature.} {\bf 433} (2005) 389-391,
  [\href{http://xxx.lanl.gov/abs/astro-ph/0501589}{{\tt astro-ph/0501589}}].

\bibitem{green}
A.~M. {Green}, S.~{Hofmann}, D.~J. {Schwarz}, 
{\it The first WIMPy halos},
{\em JCAP.} {\bf 0508} (2005) 003,
  [\href{http://xxx.lanl.gov/abs/astro-ph/0503387}{{\tt astro-ph/0503087}}].





\bibitem{slatyer}
T.~R. {Slatyer}, N.~{Padmanabhan}, D.~P. {Finkbeiner}, 
{\it CMB Constraints on WIMP Annihilation: Energy Absorption During the Recombination Epoch},
{\em Phys. Rev.} {\bf D80} (2009) 043526, 
[\href{http://xxx.lanl.gov/abs/0906.1197}{{\tt 0906.1197}}].

\bibitem{camb}
 http://camb.info/

\bibitem{cosmomc}
http://cosmologist.info/cosmomc/




\bibitem{acbar}
C.~L. {Kuo} {\em et~al.}, 
{\it Improved Measurements of the CMB Power Spectrum with ACBAR},
{\em Astrophys. J.} {\bf 664} (2007) 687,
[\href{http://xxx.lanl.gov/abs/astro-ph/0611198}{{\tt astro-ph/0611198}}].

\bibitem{boomerang}
T.~E. {Montroy} {\em et~al.},
{\it A Measurement of the CMB <EE> Spectrum from the 2003 Flight of BOOMERANG},
{\em Astrophys. J.} {\bf 647} (2006) 813,
[\href{http://xxx.lanl.gov/abs/astro-ph/0507514}{{\tt astro-ph/0507514}}].

\bibitem{cbi}
A. C.~S. {Readhead} {\em et~al.},
{\it Extended Mosaic Observations with the Cosmic Background Imager},
{\em Astrophys. J.} {\bf 609} (2004) 498,
[\href{http://xxx.lanl.gov/abs/astro-ph/0402359}{{\tt astro-ph/0402359}}].

\bibitem{vsa}
C.~{Dickinson} {\em et~al.},
{\it High sensitivity measurements of the CMB power 
spectrum with the extended Very Small Array},
{\em Mon. Not. Roy. Astron. Soc.} {\bf 353} (2004) 732,
[\href{http://xxx.lanl.gov/abs/astro-ph/0402498}{{\tt astro-ph/0402498}}].

\bibitem{0808.0881}
D.~{Cumberbatch}, M~{lattanzi}, J~{Silk}, 
{\it Signatures of clumpy dark matter in the global 21 cm background signal},
{\em Phys. Rev.} {\bf D82} (2010) 103508,
  [\href{http://xxx.lanl.gov/abs/0808.0881}{{\tt 0808.0881}}].

\bibitem{1106.1528}
S.~{Galli}, F.~{Iocco}, G.~{bertone}, A.~{Melchiorri}, 
{\it Updated CMB constraints on Dark Matter annihilation cross-sections},
{\em Phys. Rev.} {\bf D84} (2011) 027302,
  [\href{http://xxx.lanl.gov/abs/1106.1528}{{\tt 1106.1528}}].





\bibitem{fermi_cons}
{\bf Fermi-LAT} Collaboration, A.~A. Abdo {\em et~al.},
{\it Constraints on Cosmological Dark Matter Annihilation from the Fermi-LAT Isotropic Diffuse Gamma-Ray Measurement},
{\em JCAP.} {\bf 1004} (2010) 014,
  [\href{http://xxx.lanl.gov/abs/1002.4415}{{\tt 1002.4415}}].

\bibitem{pamela}
O.~{Adriani} {\em et~al.},
{\it Observation of an anomalous positron abundance in the cosmic radiation},
{\em Nature.} {\bf 458} (2009) 607,
  [\href{http://xxx.lanl.gov/abs/0810.4995}{{\tt 0810.4995}}].

\bibitem{fermi}
Fermi/LAT Collaboration,
{\it Measurement of the Cosmic Ray $e^{+}$ plus $e^{-}$ spectrum from 20 GeV to 1 TeV with the Fermi Large Area Telescope},
{\em Phys. Rev. Lett.} {\bf 102} (2009) 181101,
  [\href{http://xxx.lanl.gov/abs/0905.0025}{{\tt 0905.0025}}].


\bibitem{atic}
J.~{Chang} {\em et~al.},
{\it An excess of cosmic ray electrons at energies of 300–800 GeV},
{\em Nature.} {\bf 456} (2008) 362.

\bibitem{0909.1182}
D.~{Chowdhury}, C.~{Jog}, S.~{Vempati}, 
{\it Results from PAMELA, ATIC and FERMI: Pulsars or Dark Matter? },
{\em Pramana.} {\bf 76} (2011) 1.

\bibitem{1110.2484}
Torsten Bringmann, Pat Scott, Yashar Akrami.
{\it Improved constraints on the primordial power spectrum at small scales from ultracompact minihalos},
  [\href{http://xxx.lanl.gov/abs/1110.2484}{{\tt 1110.2484}}].

\end{thebibliography}
\providecommand{\href}[2]{#2}\begingroup\raggedright\endgroup

\end{document}